**Temperature and Humidity Dependence of Resistance in Nano-Diamond Powder**

B. de Mayo

Department of Physics, University of West Georgia, Carrollton, GA 30118
bdemayo@westga.edu

The electrical resistance of detonation nano-diamond powders was measured from liquid nitrogen temperature to room temperature and in relative humidity environments from around 10% to 100%. After sample exposures of several hours at 100% relative humidity at room temperature (around 295 K), when the temperature was reduced, the resistance increased to the upper measurement limit of our apparatus (120 MΩ) at around 240 K. Upon warming, the resistance dropped back to the room temperature value, with some hysteresis. For sample exposures after several hours at 100% relative humidity at room temperature, as the relative humidity was reduced, the sample resistance increased to the upper range limit of the apparatus. As the relative humidity was then increased (all at room temperature), the resistance dropped. For samples exposed to low (~10%) relative humidity for several hours at room temperature, as the humidity was increased (at room temperature), the resistance decreased, and then increased when the humidity was reduced. The temperature behavior was markedly differ from that of powdered graphite and multi-walled carbon nano tubes.

Introduction

Discovered in the soot of explosion experiments [1], detonation nano-diamonds have been found to have an increasing number of practical applications ranging from drug delivery to abrasion facilitators [2]. Previous work includes the measurement of the electrical resistance of nano-diamond annealed to produce graphitization and onion-like structures by Kutnetzov, et al. [3]. Kondo, et al., measured the electrical conductivity of thermally hydrogenated nano-diamond powders [4] and Mtsuko, et al., studied boron-doped nano-diamond films at low temperatures (< 5 K) [5]. The effect of water on the resistance of nano-diamond pellets is mentioned by Dolmatov [6]. Here we used a 3D printer to make rectangular sample holders and then measure the electrical resistance of detonation nano-diamond powders at temperatures between 130 K and 370 K and relative humidities at room temperature of between 10% and 100%.

Experimental

The nano-diamond samples were purchased commercially (Diamond black powder, explosion synthesized, purity: 52-65%, particle size: 4-25 nm, specific surface area: 360-420 m$^2$/g, color: black, morphology: spherical & flaky, bulk density: 0.16 g/cm$^3$, stock number 1310JGY, Nanostructured & Amorphous Materials, Inc.). Since the samples were powdered, in order to measure the electrical resistance we used a Dremel 3D printer to make rectangular sample holders out of PLA plastic, Fig. 1. Two identical sample holders were made, each was filled with nano-diamond material and slightly compressed; sample ND1 contained 0.51 ± 0.03 g and the other sample, ND2, contained 0.94 ± 0.03 g. The nano-diamond material for each was purchased separately several months apart.

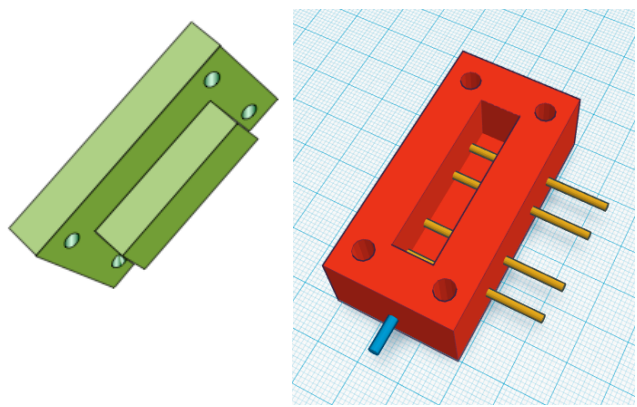

Figure 1. Sketch of a typical sample holder, constructed of PLA plastic. The overall length is 6.0 cm and the width 3.0 cm. The sample cavity is 1.0 cm wide by 1.2 cm deep by 4.0 cm. The top, left, fits into the bottom, right. The temperature sensor is shown in blue and the resistance sensing wires in orange. The outer two wires were used here; the separation distance was 3.0 cm.

The main resistance measuring system included a LabView virtual instrument incorporating three 6 1/2 digit Keithley Instruments Model 2000 digital multimeters (DMMs) connected via GPIB to the LabView computer. Two of the DMMs operated in the resistance measuring mode as controlled by the LabView virtual instrument; the stated range is from 100 μΩ to 120 MΩ. In each case, the sample temperature was determined using a resistance temperature detector (RTD) (Omega Engineering model number 1PT100KN1510) which was in physical contact with the sample. The resistance of the RTD was measured with a 4-wire connection to one of the Keithley 2000 DMMs. The resistance of the sample was measured with another one of the Keithley 2000 DMMs.

The relative humidity of the air around the sample had an effect on the sample's resistance. If the relative humidity was too low, the sample's resistance would exceed the measurement limits of the apparatus. The relative humidity in the cryostat was measured (at room temperature) with a Honeywell HIH4000-003 series channel 468 humidity sensor; its 5V power was supplied by a Hewlett-Packard HP6216A power supply. The voltage of the humidity sensor was measured using the third Keithley 2000 DMM in the DC voltage mode; stated range: from 0.1μV to 1000 V.

The cryostats used were made of expanded styrene (Styrofoam®). In some of the experiments the temperature in the cryostat was maintained with an ITC-1000F temperature controller (Inkbird Tech. Co. Ltd.) connected to a 20 W, 390 Lumen, light bulb heater. For the humidity experiments, an Incubator Warehouse Plug 'n' Play Hygrostat B122A with HumidiKit supplied a set humidity to ± 10 %. The low humidity exposure of the samples was achieved by placing them into a cryostat (volume ~ 15,000 cc) with around 100 cc of Drierite, an indicating anhydrous desiccant (W.A. Hammond Drierite Company, LTD, stock #21005, Size 4 mesh). Over time, this reduced the relative humidity environment of the sample to around 10 %.

The measurement system recorded the four parameters (the resistance of the sample and of the RTD, the voltage of the humidity sensor, and the time) at a rate of 0.5 s for runs of several hours or at a rate of up to every 10 seconds for longer runs. A 5 1/2 digit Keithley Model 2401 source measurement unit was used for the current-voltage (I-V) determination of resistance; stated ranges: 1 μV–20 V and 10 pA–1 A. This instrument was coupled to a different LabView virtual instrument via a USB-GPIB connection. Excel was used to analyze all of the data.

<u>Results.</u>

A typical plot of the temperature and resistance of nano-diamond powder as a function of time is shown in Fig. 2 for sample ND1. The sample was initially exposed to a relative humidity of around 75% for several hours. Cooling resulted in an increase in the resistance, R, until it exceeded the upper range limit of our apparatus, 120 MΩ, at 246 K. Upon warming, R returned in-range at about 248 K. These results are plotted as R vs. temperature, in Fig. 3; the best fit was to a 6th degree polynomial:

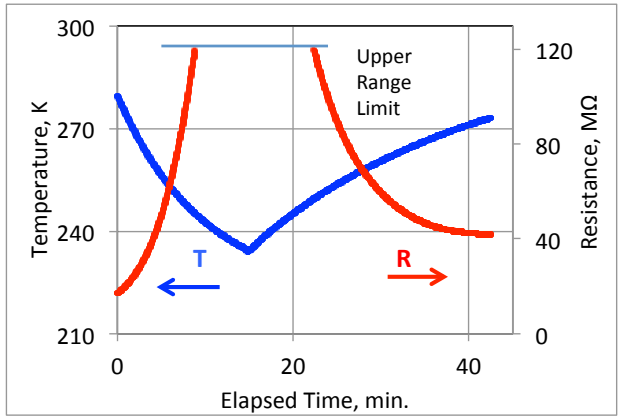

Figure 2. Temperature, T, blue, and resistance, R, red, vs. elapsed time, t. The upper range limit is about 120 MΩ, reached at 246 K upon cooling and 248 K upon warming. Initial relative humidity was around 75%.

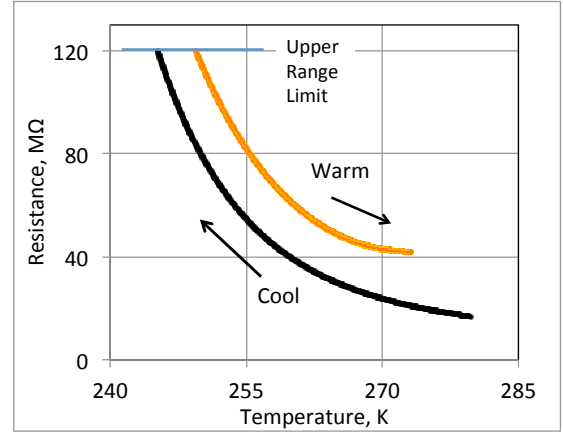

Figure 3. Resistance, R, vs. temperature, T, sample ND 1. Cooled first (black) then warmed (orange).

$R(T) = (6.990 \times 10^{-8} \ \Omega/K^6) \times T^6 - (1.124 \times 10^{-4} \ \Omega/K^5) \times T^5 + (7.535 \times 10^{-2} \ \Omega/K^4) \times T^4 - (2.696 \times 10^{1} \ \Omega/K^3) \times T^3 + (5.429 \times 10^3 \ \Omega/K^2) \times T^2 - (5.837 \times 10^5 \ \Omega/K) \times T + (2.617 \times 10^7 \ \Omega),$

$R^2 = 1.000$, cooling,

$R(T) = -(1.485 \times 10^{-7} \ \Omega/K^6) \times T^6 + (2.281 \times 10^{-4} \ \Omega/K^5) \times T^5 - (1.458 \times 10^{-1} \ \Omega/K^4) \times T^4 + (4.962 \times 10 \ \Omega/K^3) \times T^3 - (9.487 \times 10^3 \ \Omega/K^2) \times T^2 + (9.657 \times 10^5 \ \Omega/K) \times T - (4.088 \times 10^7 \ \Omega),$

$R^2 = 1.000$, warming,

where $R^2$ is the Microsoft Excel coefficient of determination and T is in units of K. The initial relative humidity was about 75%.

Fig. 4 and Fig. 5 show the effect at room temperature (around 295 K) of relative humidity and time on the resistance of sample ND 2 as the relative humidity drops from around 75% to 14%. The resistance vs. time curve of Fig. 4 could be fit well to a cubic polynomial:

$R(T) = + (2.276 \ \Omega/K^3) \times T^3 + (3.317 \ \Omega/K^2) \times T^2 + (3.659 \ \Omega/K) \times T + (31.45 \ \Omega),$

$R^2 = 0.99984$, $R^2$ and T as before.

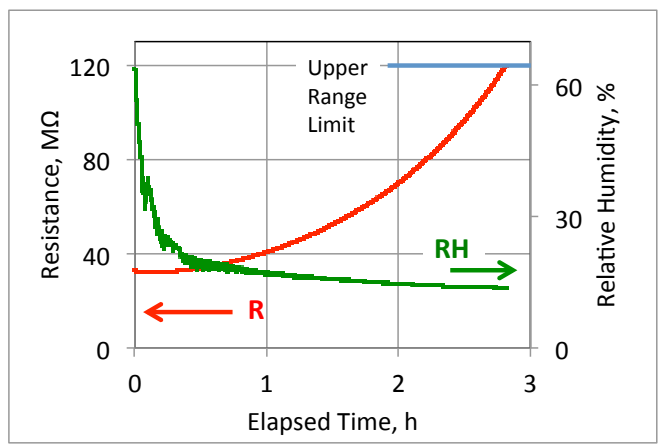

Figure 4. Resistance and relative humidity vs. elapsed time for nano-diamond ND2 at room temperature.

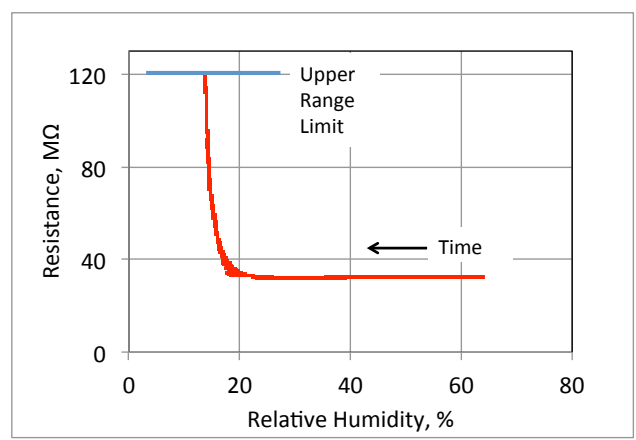

Figure 5. Resistance vs. relative humidity for nano-diamond ND2 at room temperature.

Fig. 6 shows the results over a period of 24 hours for the resistance of sample ND2 at room temperature as the relative humidity was increased from 37% to 54% and dropped to 47%. The resistance was greater than the upper range limit of 120 MΩ until the humidity had reached around 53% at the 5 hour mark. After 8 more hours, the resistance kept dropping steadily until that point where the relative humidity started dropping, whereupon the resistance rose steadily, reaching the upper range limit again after 11 hours and 47 % relative humidity. The resistance was relatively steady for the 2 hours around the 15 hour elapsed time mark.

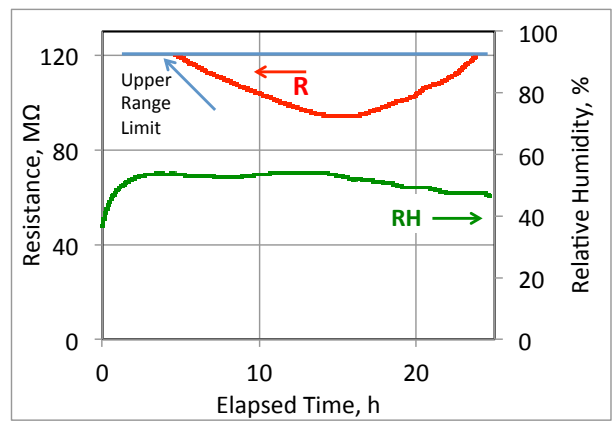

Figure 6. Resistance (red) and relative humidity (green) vs. elapsed time for sample ND2 at room temperature.

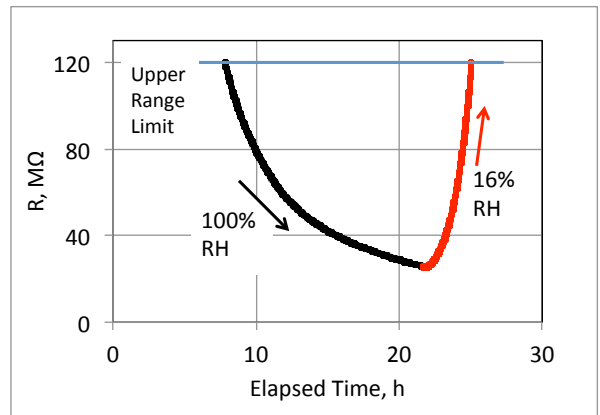

Figure 7. Resistance vs. elapsed time at room temperature for first exposure to 100% relative humidity (black) and then at 22 hours to 16% relative humidity, (red); sample ND2.

Fig. 7 shows the resistance of sample ND 2 at room temperature as the cryostat first goes from a relative humidity of 47% to 100% relative humidity in about 10 minutes. After 8 hours, the resistance drops into the measurable range at 120 MΩ. At 22 hours elapsed time, the resistance is has fallen to around 27 MΩ, whereupon the relative humidity in the cryostat is reduced, reaching 25% in 30 minutes and 16% in 2 hours (22 hours elapsed time). The resistance increases to the upper range limit at 25 hours elapsed time. The response to low humidity seems faster than to high humidity.

In Fig. 8, we see the time dependence of the resistance of sample ND2 when it goes from a low humidity environment (~10%) to environments of about 100% and of about 45% relative humidity. The resistance in the case of exposure to 100% relative humidity drops to the upper range limit of 120 MΩ faster than the 45% case: 3.2 hours vs. 4.7 hours, respectively, and reaches a lower final limit at 20 hours of (4.70 ± 0.5) MΩ vs. (8.17 ± 0.4) MΩ (errors estimated), respectively. The two different samples were in close agreement in all of the tests, as can be seen typically in Fig. 9. The samples were purchased one year apart from the same supplier. The best least-squares fit for all of the curves of Fig. 7, 8, and 9 were 6th degree polynomials; the lowest $R^2$ was 0.998.

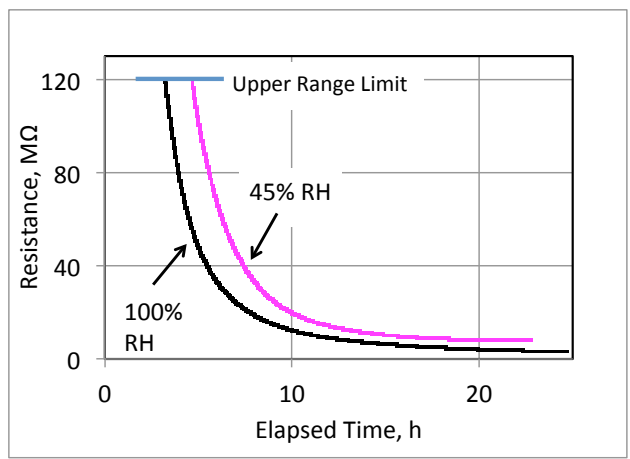

Figure 8. Resistance vs. elapsed time for sample ND2 at room temperature, previously exposed to about 10% relative humidity and then (here) to a relative humidity of about 100% (black) and to a relative humidity of about 45% (magenta).

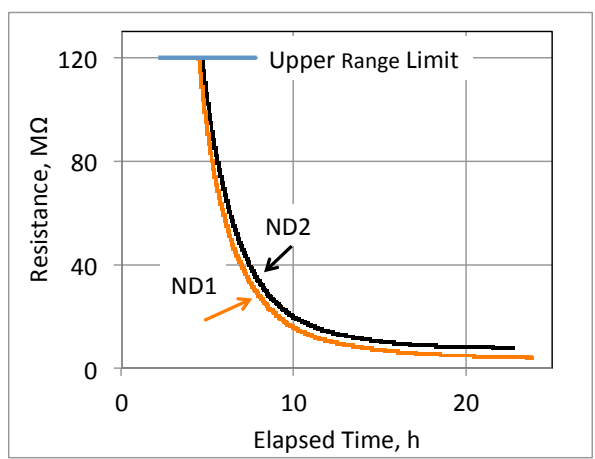

Figure 9. Resistance vs. elapsed time for samples ND1 (orange) and ND2 (black) at room temperature. The samples were previously exposed to about 10% relative humidity and then (here) to about 100% relative humidity.

With the same set-up and identical sample holders, we measured the temperature dependence of the resistance of graphite (lab grade, powdered, dry, Aldor Corp., #9475566, CAS # 7782-42-5) and a sample of carbon multi-wall nanotubes (Part No. MWNT-12950040-00, Lot No. BMCD05210006, Helix Material Solutions) at about 70% relative humidity at room temperature. Fig. 10 shows the results, which can be likened to the nano-diamond results of Fig. 3.

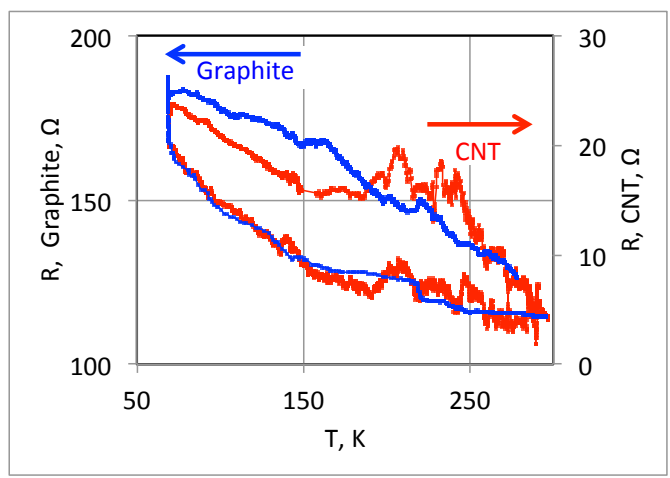

Figure 10. Resistance, R, vs. temperature, T, for powdered graphite (blue) and carbon nanotubes (CNT, red). The samples were first cooled from room temperature, 295 K, to 70 K temperature and then warmed back to room temperature.

At room temperature, for nano-diamond sample ND1 at about 70% relative humidity, and a resistance of 4.80 MΩ measured with our regular apparatus, a Keithley 2401 source meter was used to obtain the current vs. voltage plot given in Fig. 11. The almost-linear data are best fitted to a quadratic:

$$V(I) = 0.0406 \text{ x } I^2 \text{ (V/A}^2) + 2.639 \text{ x I (V/A)} + 0.0007 \text{ V,}$$

I in amps and $R^2 = 0.9992$, indicating an electrical resistance of $(2.64 \pm 0.01)$ MΩ at I = 0. The error is estimated. A similar plot for sample ND2, resistance of 11.36 MΩ, also measured with our regular apparatus, yields a relation:

$$V(I) = 0.0660 \text{ x } I^2 \text{ (V/A}^2) + 8.220 \text{ x I (V/A)} - 0.00567 \text{ V,}$$

with I in amps and with an $R^2$ of 0.9988. This indicates an electrical resistance of $(8.22 \pm 0.01)$ MΩ at I = 0. This error is estimated, too.

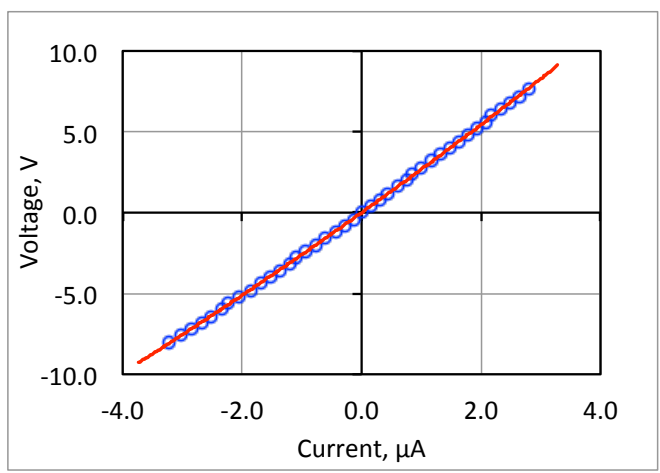

Figure 11. Voltage vs. current for nano-diamond sample ND1 at room temperature and relative humidity of about 70%. A quadratic fit (red) gives the resistance at I=0 to be $(2.64 \pm 0.01)$ MΩ (error estimated).

Our results can be compared to those of Dolmatov [6], who found a resistivity for compressed nano-diamond pellets of 10 - 100 MΩm (dry), "After moistening the pellet, the resistivity sharply increases to ($< 10^3$ Ωm for a sample with 5%water)," [6]. At 302 K and in 84 % relative humidity, the resistance of sample ND1 was $(109.05 \pm 5)$ MΩ and its resistivity was $(110 \pm 11)$ MΩm, in somewhat agreement with Dolmatov [6]. However, we found that increasing the humidity of the air surrounding a sample <u>reduced</u> its resistance and thus its resistivity, Fig. 4-9.

We observed no aging effects over a period of 2 years. The two samples behaved similarly in all measurements.

Conclusions

The results indicate that temperature and water, in the form of the relative humidity surrounding the sample, play unknown but crucial roles in the electrical resistance of nano-diamonds.

Acknowledgements:
        The author gratefully acknowledges the generous support of the Georgia Space Grant Consortium-NASA NNG05GJ65H and the excellent assistance of undergraduate research workers Hannah Watkins and Katlyn Brumbelow.